# Soft matter and fractional mathematics: insights into mesoscopic quantum and time-space structures

Wen Chen[*]
Computational Physics Laboratory, Institute of Applied Physics and Computational Mathematics, P.O. Box 8009, Beijing 100088, P. R. China
(29 April 2004)

**Abstract**: Recent years have witnessed a great research boom in soft matter physics. By now, most advances, however, are of empirical results or purely mathematical extensions. The major obstacle is lacking of insights into fundamental physical laws underlying fractal mesostructures of soft matter. This study will use fractional mathematics, which consists of fractal, fractional calculus, fractional Brownian motion, and Levy stable distribution, to examine mesoscopic quantum mechanics and time-space structures governing "anomalous" behaviors of soft matter. Our major new results include fractional Planck quantum energy relationship, fractional phonon, and time-space scaling transform.

## 1. Introduction

The physical principles apply on the physical size scale. Einstein's relativity theory has revealed our universe on the large scale and shown a profound link between time and space. But the theory does not hold at the microscopic subatomic level, where quantum mechanics theory prevails. Between the macroscopic and microscopic levels is the mesoscopic level, in which scope lie the functioning elements of soft matter, also known as soft condensed matter or complicated fluids. Among the best-known soft matter are polymers, colloids, emulsions, foams, living organisms, rubber, oil, soil, other porous media, etc. The behaviors of soft matter have long been observed "bizarre" or "anomalous" compared with those of the ideal solid and fluid matter and can not arise from relativistic or quantum properties of elementary molecules [1]. The current consensus is that the fractal mesostructures of soft matter dramatically alter the physical properties in new ways. This prompts us to consider the new physical principles governing these materials whose fractal mesostructures dominate their physical behaviors.

The mesoscopic scale property means that the large amount of the elementary molecules is grouped together on mesoscopic scale, and behaves like a macromolecule, and the mesostructures and interactions of these bonded groups determine macroscopic behaviors. Soft matter has both solid and fluid properties in the sense that the elementary molecules within the fastened-together-molecule group are constrained like those of an ideal solid, while each group moves like the molecules of a simple fluid. The focus of much research has by now been on fractal, periodic, or dilation symmetric mesostructures, self-organization, and cooperative organized motion of soft matter. Fractional calculus and fractal have been found in recent years to be a useful tool in the mathematical modeling of physical behaviors of soft matter. The traditional endeavor, however, is mostly to fit the observed findings and lacks an in-depth investigation into the basic physics, and a variety of empirical or semi-empirical phenomenological constitutive model equations have emerged. On the other hand, some pure mathematical extensions also have recently been proposed through a simple replacement of some time/space derivative terms in the existing model equations by the fractional derivatives but are unfounded on the real-world observations. In this study, based on our recent phenomenological model of acoustic dissipation of soft matter, the purpose of this note is to explore further into the fundamental physical principles. The fractional calculus, fractal, Levy stable distribution, 1/f noise, Hurst exponent, and fractional Brownian motion are all inherently related, and we thus call these mathematical tools the fractional mathematics, which are all indispensable for a systematic study of soft matter.

---
[*] Email: chen_wen@iapcm.ac.cn

In what follows, we start our analysis in section 2 from a particular case: anomalous dissipation of the acoustic wave propagating through soft matter. The recent phenomenological model equation of this process via the fractional Laplacian [2] is analyzed to reveal the inherent fractional quantum energy band of soft matter. Then we introduce the fractional Planck quantum energy relationship and the novel concept of the fractional phonon. The energy distribution of macromolecules of soft matter is also derived. In section 3, based on the fractional Planck quantum energy, we derive the so-called fractional Schrödinger's equation (FSE) and analyze its dissipative property. The fractional Riesz potential is also included in the present FSE. And finally, section 4 presents the time-space scaling transform (deformation) of the fractal mesostructures of soft matter with the help of the fractional Brownian motion and the Levy stable distribution.

## 2. Anomalous dissipation and fractional quantum

The frequency-dependent energy dissipation of acoustic wave propagation through soft matter can be expressed as $E = E_0 e^{-\alpha(\omega)z}$, where $E$ represents the amplitude of an acoustic field variable such as pressure, $z$ is the traveling distance, and $\omega$ denotes angular frequency [3]. The attenuation coefficient $\alpha(\omega)$ for a wide range of frequencies of practical interest is found to obey an empirical power law function of frequency [4]

$$\alpha(\omega) = \alpha_0 \omega^y, \qquad y \in [0,2], \qquad (1)$$

where $\alpha_0$ and $y$ are media-specific parameters obtained through a fitting of measured data[1,2]. For most solid and highly viscous materials, $y$ is close to 2; while for most soft matter such as biomaterials, sediments, and rock layers, $y$ is a real number, ranging from 1 to 1.7. For $y \neq 0,2$, the dissipation process can not be described by common time-space partial differential operators of integer order and is thus called anomalous dissipation.

Among all existing modeling methodologies, the fractional derivative approach requires the fewest number of experimentally ready parameters in the phenomenological constitutive modeling of this fractional power law dissipation process [2,3]. Instead of the traditional fractional time derivative, the present author recently used the fractional Laplacian in space, also known as the fractional Riesz derivative, to develop a mathematical model

$$\Delta^2 s = \frac{1}{c_0^2}\frac{\partial^2 s}{\partial t^2} + \frac{2\alpha_0}{c_0^{1-y}}\frac{\partial}{\partial t}(-\Delta)^{y/2} s, \qquad (2)$$

where $c_0$ is the small signal sound speed, $s$ denotes the pressure, and $(-\Delta)^{y/2}$ represents the fractional Laplace, which is a positive definite operator [2,5]. By using the time-space Fourier transform, equation (2) was found to coincide with the frequency power law dissipation (1). It is worth pointing out that equation (2) concerns with a phenomenological time-space representation rather than a theoretical foundation for the empirical frequency power law dissipation (1). In other words, the model is to describe the macroscopic dissipation phenomena but do not necessarily reflect physical mechanisms behind the scenes. The purpose of this study is to unravel the fundamental physical mechanism underpinning this model.

Assuming the left-side term of equation (2) has little impact and ignoring it, and then integrating with respect to time $t$, we have the well-known anomalous diffusion equation

$$\frac{\partial s}{\partial t} + \gamma(-\Delta)^{y/2} s = 0, \qquad (3)$$

where $\gamma = 2\alpha_0 c_0^{1+y}$. By using the separation of variables, namely, $s(x,t) = T(t)Q(x)$, we have

$$\frac{dT}{dt} + \lambda^2 \gamma T = 0, \qquad (4)$$

$$(-\Delta)^{y/2} Q - \lambda^2 Q = 0, \qquad (5)$$

where $\lambda$ is a positive constant since the fractional Laplacian is a positive operator. Eq. (5) is a fractional Helmholtz eigenvalue equation. Its eigenfunction in an unbounded domain is $\exp\{ipx\}$, where $\lambda^2 = |p|^y$. It is well know that the eigenvalues of the ordinary Laplacian are minima of the potential energy [6]. It is straightforward to extend this theorem to the eigenvalues of the fractional Laplacian. Now we see the discrete potential energy spectrum of fractional order in mesoscopic macro-molecule of soft matter. This implies that unlike ideal solids such as crystals, soft matter has the fractional quantum energy band. The observation also provides us a key hint leading to a fractional quantum energy relationship conjecture

$$E = \hat{h}\omega^{y/2} \tag{6}$$

to explain the fractional power dependence of energy dissipation of soft matter on frequency, where $\hat{h}$ is the scaled Planck constant. When $y=2$, (6) is reduced to the normal Planck quantum energy relationship. Laskin [7] also derives a fractional energy spectrum of an imaginary hydrogenlike atom, called the fractional "Bohr atom", in terms of his fractional Schrodinger equation, which will be discussed in the next section 3. But [7] does not propose (6).

The acoustic wave propagation is actually vibration of the molecules of media, and the quantized energy of this vibration (quantum oscillator) is called phonons. The wave propagation is the motion of phonons (acoustic wave packets) through the material. Although the phonon is defined as a particle of energy in the movement of molecules [8], it has the wave nature (wave-particle duality). Traditionally, the phonon is mostly involved in the study of the atomic lattice of the crystal, being the quanta of lattice vibration energy, and is intimately related to the physical structures of materials. However, there are no lattice structures (at most fractal lattice) in the mesoscopic macromolecules of soft matter (e.g., porous media and amorphous solids), and it is thus difficult to apply the phonon concept here. It is seen from (6) that acoustic energy is transmitted and absorbed in a fractional quantum unit $\hat{h}\omega^{y/2}$ through soft matter. This suggests that the acoustic wave energy actually exists in the fractional quantum phonon form while traveling through soft matter. This fractional phonon underlies the fundamental physical mechanism of the fractional frequency-dependent power-law dissipation. Since the phonon concept is essential in solid state physics, responsible for many physical properties of materials, the concept of the fractional phonon will be useful in the analysis of thermal and electrical conductivity of soft matter (thermal and optical phonons).

The one-dimension anomalous diffusion equation (3) can be characterized by the so-called Levy stable distribution, whose Fourier transform has the form [9-11]

$$P(k) = e^{-\gamma |k|^y}, \tag{7}$$

where $k$ is the wavenumber, and the absolute value is its module. In this study, the parameter $y$ is considered the stability index of the Levy distribution. Extending the previous fractional Plank quantum energy relationship to the momentum, we have

$$\vec{p} = \hat{h}\vec{k}^{y/2}, \tag{8}$$

where $p$ denotes the momentum. Since kinetic energy $E_k = p^2/2m$, substituting (8) into (7) produces

$$P(E_k) = e^{-2m\gamma E_k/\hat{h}}. \tag{9}$$

The Maxwell-Boltzmann distribution (9) is the possibility of macromolecules of soft matter in terms of energy state. The fractional phonons obey the Bose-Einstein statistics.

**3. Fractional Schrödinger's equation**

On mathematic aesthetic (12) or physical heuristic (7,13,14) grounds, the fractional Schrodinger equations have recently been proposed to use fractional time and space derivatives instead of the integer-order derivatives in the standard Schrodinger equation. These FSEs are justified somewhat a posteriori on the ground they solve some problems or derive some reasonable results. None of them, however, is a consequence of a basic principle of physics. Quantization of a fractional-derivative Hamiltonian can not be achieved by the traditional replacement of momenta with coordinate derivatives. This section will derive the FSE via the fractional Planck energy relationship hypothesis presented in section 2.

Broadly speaking, there are two types of the modeling methodology, space and time operations to describe the anomalous dissipation of acoustic wave propagation through soft matter. The above-mentioned space fractional Laplacian model (2) reflects the spatial fractal irregularity, while the time thermoviscous operator underlies the memory effect, also often called heredity, relaxation or hysteresis in some publications, which depends on the past history of motion. A mixed space-time model may be appealing to be the alternative third type of models. If both the fractional time and space derivatives are used, the dissipative wave equation (2) will be restated as [15]

$$\Delta s = \frac{1}{c_0^2}\frac{\partial^2 s}{\partial t^2} + \varepsilon \frac{\partial^\eta}{\partial t^\eta}(-\Delta)^{\tau/2} s, \qquad (10)$$

where $0<\eta<2$ is real valued, $\eta + \tau - 1 = y$, and $\varepsilon$ is the viscous coefficient. Repeating the same process to reduce (2) to (3), (10) can be reduced to a fractional diffusion-wave equation

$$\frac{\partial^\alpha s}{\partial t^\alpha} = -\kappa(-\Delta)^{\tau/2} s. \qquad (11)$$

where $\alpha = 2 - \eta$, $\kappa = c_0^2 \varepsilon$. If we replace $-\kappa$ by a purely imaginary $e^{-i\pi\alpha/2}\hat{h}/2m$ and add an external potential force $V(\vec{r})s$ [16], we get the fractional Schrodinger equation

$$e^{i\pi\alpha/2}\hat{h}\frac{\partial^\alpha s}{\partial t^\alpha} = \frac{\hat{h}^2}{2m}(-\Delta)^{\tau/2} s + V(\vec{r})s. \qquad (12)$$

The above approach to derive the fractional Schrodinger equation is not physically rigorous. Now we try to do so via the fractional Planck energy relationship (6) and (8). Consider the quantum plane wave $\Psi = Ae^{i\vec{k}\vec{r}-i\omega t}$, we have

$$(-\Delta)^{y/2}\Psi = |k|^y \Psi = \frac{p^2}{\hat{h}^2}\Psi. \tag{13}$$

Thus,

$$E_k = \frac{p^2}{2m} = (-\Delta)^{y/2}\frac{\hat{h}^2}{2m}. \tag{14}$$

On the other hand,

$$\frac{\partial^{y/2}\Psi}{\partial t^{y/2}} = (-i\omega)^{y/2}\Psi = e^{-i\pi y/4}\frac{E}{\hat{h}}\Psi \rightarrow e^{i\pi y/4}\hat{h}\frac{\partial^{y/2}\Psi}{\partial t^{y/2}} = E\Psi \tag{15}$$

The classical Hamiltonian for a particle with potential energy $V(\vec{r})$ reads

$$E = E_k + E_p = \frac{p^2}{2m} + V(\vec{r}). \tag{16}$$

Replacing all dynamic variables in (16) with the equivalent operator, the fractional quantum mechanical Hamiltonian can be stated as

$$e^{i\pi y/4}\hat{h}\frac{\partial^{y/2}\Psi}{\partial t^{y/2}} = \frac{\hat{h}^2}{2m}(-\Delta)^{y/2}\Psi + V(\vec{r})\Psi. \tag{17}$$

(17) is found to be a special case of (12) in that the orders of the temporal and spatial derivatives are identical in the former. Unlike the traditional conservative quantum Hamiltonian to describe a particle moving in a conservative field, the present one is dissipative to reflect the dissipative mesoscopic field of soft matter thanks to the presence of the fractional derivative. The differentiability constraints on wavefunction $\Psi$ is also relaxed, and it only needs to be differentially continuous on the fractional order.

In fact, the present fractional Schrodinger equation underlies the two basic regimes leading to anomalous dissipative (diffusion) process, namely, Levy process (Levy flight, long tailed distribution) and fractal time (fractional Brownian motion, long-range correlation), which cause partially time irreversibility and ergodicity breaking. These two mechanisms produce strikingly common characteristic process, such as the power-law growth of the second moment, but are fundamentally different and in fact the cooperative physical paradigms to describe complicated phenomena arising from fractal time-space mesostructures of macromolecules of soft matter.

The potential energy results from position or configuration associated with the intermolecular attractive forces. A conservative potential energy of an object is dependent upon its position and not upon the path by which it reached that position. Soft matter has been observed clear memory properties and path dependence, and thus its potential filed is not conservative. However, the potential energy in the FSE proposed in [7,14-16] is still the traditional integer order, implying a conservative potential field. In terms of the fractional Riesz potential concept, we have a full fractional FSE

$$e^{i\pi y/4}\hat{h}\frac{\partial^{y/2}\Psi}{\partial t^{y/2}} = \frac{\hat{h}^2}{2m}(-\Delta)^{y/2}\Psi + WV_d^y(\vec{r})\Psi. \tag{18}$$

where $V_d^y$ is the $y$-order Riesz potential under topological dimension $d$, and $W$ is the strength of the potential. Serving as an illustrating example, below is the equation for one-dimensional quantum harmonic oscillator subject to a fractional potential

$$i\hbar\frac{\partial\Psi}{\partial t} = -\frac{\hbar^2}{2m}\frac{\partial^2\Psi}{\partial x^2} + Qx^y\Psi, \tag{19}$$

whose solution has the non-Gaussian form (Levy distribution) $e^{-\alpha x^y}$. [7] also gives a fractional oscillator having the similar form of (19) in the potential energy term.

In summary, the mesostructures of soft matter play an essential role in determining the exponent $y$ of the frequency power law (1). And $y$ is believed to be the trial fractal, representing the irregular boundary of the fractal macromolecules of soft matter. And acoustic energy propagating through soft matter is transmitted and absorbed in terms of fractional quantum energy band decided by the molecule boundary fractal. These macromolecules undergo the nonlocal visco-elastic interactions (not a perfectly elastic collision) to transfer the kinetic and potential energies [17].

## 4. Mesoscopic time-space structures

The fractional wave-diffusion equation (11) in one dimension underlies a deviation from Gaussian scaling of the mean square displacement dependence on time [11]

$$\langle(\Delta x)^2\rangle \propto t^{2\alpha/\tau}. \tag{20}$$

When $\alpha=1$, $\tau=2$, we have the normal Gaussian scaling. It is possible to derive the fractional Planck energy relationship through a Fourier transform of (20) and the corresponding velocity autocorrelation function. When $\alpha=1$, $\tau<2$, (20) leads to the diverging moment of higher order

$$\langle(\Delta x)^n\rangle = \infty, \quad \tau < n. \tag{21}$$

Thus, for $2\alpha/\tau \neq 1$, the mean square displacement diverges, which implies the potential energy can not trap the particle and the velocity correlation has a non-exponential tail. Following this view, the mean kinetic energy for a finite mass $m$ also diverges [11],

$$\left\langle \frac{1}{2}mv^2 \right\rangle = \infty, \; \alpha = 1, \tau < 2. \tag{21}$$

By analogy with the time dilation and length contraction in the Einstein's relativity theory, the novel approach to solve the present paradox is to define the new observation scales of fractal mesoscopic time and space. Namely,

$$\begin{cases} x' = x^{\tau/2}, \\ t' = t^{\alpha}. \end{cases} \tag{22}$$

Then, by using (22), the anomalous scaling of the mean square displacement (20) is restored to the normal Gaussian

$$\left\langle (\Delta x')^2 \right\rangle \propto t', \tag{23}$$

whose the second moment is finite, and the corresponding mean kinetic energy exists.

The time and space scaling transform (22) reflects the fractal time-space mesostructures of macromolecules of soft matter. The Feynman integral over such a scaled time-space path leads to the fractional quantum mechanics [7]. The fractional Planck energy relationship (6) is established under such scaled time-space mesostructures and characterizes the energy fluctuations with a Levy process in space and a Brownian motion in time.

Corresponding the time-space scaling transform, we propose a new definition of the fractional derivative

$$\frac{d^{\tau} x}{dt^{\vartheta}} == \lim_{\Delta t \to 0} \frac{\Delta x^{\tau}}{\Delta t^{\vartheta}}. \tag{24}$$

Unlike the existing fractional derivatives, the upper and lower indices of the derivative can be different in (24). We call it the generalized fractional derivative. The different upper and lower indices may be feasible for the Caputo derivative and fractional Laplacian. In terms of (24), we define a generalized velocity

$$\widetilde{v} = \frac{d^{\tau/2} x}{dt^{\alpha}}. \tag{25}$$

Then, we further postulate a plausible kinetic energy definition in the fractal soft matter

$$\widetilde{E}_k = \frac{1}{2} m \widetilde{v}^2. \tag{26}$$

The kinetic energy $\widetilde{E}_k$ corresponds to the scaled time–space (22) of fractal macromolecules and exists in soft matter.

Einstein's generalized relativity defines mass through gravitation as the manifestation of the time-space structures. Following this view, it is reasonable to introduce the fractal mass modified by the fractal time-space mesostructures

$$\widetilde{m} = m^\nu, \tag{27}$$

where $\nu$ is the scale parameter, a function of fractal time-space scales $\alpha$ and $y$. We will discuss this issue in a subsequent paper separately.

(22), (9), (23), and (26) are in fact to reflect the scale invariance of classical physical laws, in particular, for adapting them to describe the physical behaviors of the fractal soft matter.

Another issue of concern in this section is the complex velocity. For the one-dimension dissipative quantum plane wave $\Psi = Ae^{i k \bar{r} - i\omega t}$, $k = k_r + i k_i$, where the imaginary part $k_i = \alpha_0 \omega^y$ is responsible for dissipation. The fractional Planck relationship (8) leads to

$$p = mv = \hat{h} k^{y/2}. \tag{28}$$

Then

$$p = p_r + i p_i = m(v_r + v_i). \tag{29}$$

The complex velocity occurs in (29). Otherwise, we have to introduce the complex mass concept ("dark matter"?) to accommodate the complex momentum. The imaginary velocity corresponds to the dissipation dictated by the geometry of fractal time-space mesostructures of soft matter. The complex fractional derivative introduced in [18] can be used to evaluate the complex velocity and, moreover, may be an important mathematical tool to modify the Schrödinger's equation for the inclusion of the complex potential (a further extension of the fractional Riesz potential). Nottale [19] also proposes the scale relativity theory to explain such anomalous phenomena, where the complex velocity is also mentioned from the mathematical non-differential point of view. It is also noted that the generalized fractional derivative presented here is also different from the scale derivative proposed in [19].

Before ending this section, we emphasize that we do not have a rigorous mathematical basis for the above daring explorations.

## 5. Concluding remarks

Among the major new findings of this study are the fractional Planck energy relationship, fractional phonon and energy band, time-space scaling transform, generalized velocity and kinetic energy, complex velocity. The standard quantum theory is found to correspond statistically to the Gaussian distribution (space) and Brownian process (time), while the present fractional quantum models underlie the fractional Brownian motion in time and the Lévy stable distribution in space. We also introduce the generalized fractional derivative to meet our modeling demand.

In essence, the fractional mathematics is the mathematical devices to depict the history-dependence and global interaction of systems. As pointed out by Sir Atiyah [20], "Perhaps we need to know the past in order to predict the future. Perhaps the universe has memory." In recent decades, the fractional calculus has been realized to be an indispensable tool in phenomenally describing "bizarre" behaviors of soft matter. This study shows that the fractional mathematics can also lead us to some of the deepest portions of the basic physical principles in the fractal mesoscopic world. As Baglegy and Torvik [21] put it, the fractional calculus equation representations "should be viewed as something more than an arbitrary construction which happens to be convenient for the description of experimental data". To some extent, we believe that the fractional quantum mechanics and time-space scaling transform are the underlying mechanism of a large number of diverse complicated systems which are now phenomenally modeled from varied viewpoints by a variety of the fractional mathematic methodology, e.g., 1/f noise, fractional calculus, fractal, Levy process, Hurst exponent, fractional Brownian motion.